\documentclass[journal]{rmaa}

\usepackage{paralist}
\def\RRab{RRab}
\def\RRc{RRc}
\usepackage{psfrag,color}

\usepackage[T1]{fontenc}
\usepackage[utf8]{inputenc}

\title{A new visit to the variable stars in M56 and its Colour Magnitude diagram structure} 

\author{
  D. Deras,\altaffilmark{1} 
  A. Arellano Ferro,\altaffilmark{1}
  I. Bustos Fierro,\altaffilmark{2}
  and M. A. Yepez\altaffilmark{1}}

\altaffiltext{1}{Instituto de Astronom\'ia, Universidad Nacional Aut\'onoma de M\'exico, Ciudad Universitaria, C.P. 04510, M\'exico.}

\altaffiltext{2}{Observatorio Astron\'{o}mico, Universidad Nacional de
C\'{o}rdoba,  Laprida~854, X5000BGR, C\'{o}rdoba, Argentina.}

\shortauthor{Deras, Arellano Ferro, Bustos Fierro \& Yepez }
\shorttitle{VARIABLE STARS IN M56 AND ITS COLOUR MAGNITUDE DIAGRAM}

\abstract{We present a $\emph{VI}$ CCD photometric study and a membership analysis of the globular cluster M56 (NGC 6779). This produced a decontaminated CMD from field stars, which enabled a better confrontation with theoretical isochrones; Zero-Age Horizontal Branches (ZAHB) and post-ZAHB evolutionary tracks. Post He-flash evolutionary models with a He-core mass of 0.5 M$_\odot$ and envelopes of 0.04 - 0.18 M$_\odot$, cover the complete Horizontal Branch. Models with total mass $\sim$ 0.68 M$_\odot$ explain the RR Lyrae, while those with a mass $\sim$ 0.56 M$_\odot$ and a very subtle envelope, explain the Pop II cepheids with a progenitor in the blue tail of the HB. 
Based on the Fourier decomposition of the $V$ light curve of a single cluster member RRc star, we determined a metallicity of $\rm [Fe/H]_{ ZW}$ = -1.96 $\pm$ 0.09. Several independent distance determination approaches lead to a mean distance to M56 of $\big< d \big>$ = 9.4 $\pm$ 0.4 kpc. Finally, we report 5 new variables: 1 SX Phe, 3 EB, and 1 RRc.}
\resumen{Presentamos un estudio fotom\'etrico y un an\'alisis de membres\'ia del c\'umulo globular M56 (NGC 6779) con im\'agenes CCD $\emph{VI}$. Lo anterior permiti\'o construir un diagrama color-magnitud sin contaminaci\'on de estrellas de campo, que facilita la comparaci\'on con modelos de is\'ocronas, rama horizontal de edad cero (ZAHB) y trazas evolutivas post-ZAHB. Modelos con masas de 0.5 M$_\odot$ y envolventes de 0.04 - 0.18 M$_\odot$, representan toda la rama horizontal (HB). Modelos con masas $\sim$ 0.68 M$_\odot$ son adecuados para estrellas RR Lyrae, mientras que masas de $\sim$ 0.56 M$_\odot$ y envolventes muy delgadas explican a las estrellas de Pob II con un progenitor en la cola azul de la HB. 
La descomposicion de Fourier de la curva de luz de la \'unica RRc miembro del c\'umulo sugiere la metalicidad $\rm [Fe/H]_{ ZW}$ = -1.96 $\pm$ 0.09. La distancia media de M56 obtenida por diversos m\'etodos independientes es $\big< d \big>$ = 9.4 $\pm$ 0.4 kpc. Reportamos 5 nuevas variables: 1 SX Phe, 3 EB, y 1 RRc.}

\addkeyword{Stars: fundamental parameters}
\addkeyword{Stars: variables: RR Lyrae}
\addkeyword{Globular Clusters: Individual: M56}

\begin{document}
\maketitle

\section{Introduction} \label{intro}

The globular cluster M56 (NGC 6779) is among the most metal-poor clusters in the Galaxy, is a relatively nearby and hence a bright system lying well within the Galactic disc ([Fe/H]$\sim$ -1.98 \citep{Searle1978}; $d$=8.4 kpc \citep{Hatzidimitriou2004}; ($l = 62.66^{\circ}$, $b = 8.34^{\circ}$), therefore, it is subject to a considerable reddening ($E(B-V)=0.26$) \citep{Kron1976}. Typical of very metal-poor clusters, its Horizontal Branch (HB) shows a very long and populated blue tail and a scarcely populated instability strip (IS). As a consequence, very few RR Lyrae stars are expected to be found in the system. In fact, presently only 2 RRab and 2 RRc are listed in the Catalogue of variable stars in globular clusters (CVSGC) of \citet{Clement2001}. Other variables included in the CVSGC are 1 W Vir (CW), 6 semi-regular or slow variables (SR or L), 1 RV Tauri (RV) and 1 SX Phoenicis star, for a total of 14 variables. The variability of V2 has been doubted by several authors and we shall report our conclusion in the present paper. The most recent light curve analysis of the known variables is by \citet{Pietrukowicz2008}, which shall be of substantial basis for our present analysis. It has been argued, on kinematical grounds, that  M56 might be of extragalactic origin due to the occurrence of a massive accretion event by the Milky Way, around 10 Gyrs ago \citep{Deason2013}. A study conducted by \citet{Piatti2019} concluded that M56 shows evidence of extra-tidal features like tails and extended haloes, as a consequence of the merger event. Also,  \citet{Massari2019} showed that the Age-Metallicity relation for Galactic globular clusters is bifurcated, meaning that metal-rich clusters with disc-like kinematics are more likely to have been formed \textit{in situ}, while metal-poor clusters with halo-like kinematics are more likely to have been accreted, M56 being part of the latter.

In the present study we propose the determination of the cluster mean metallicity and distance, as indicated by the light curve morphology of its RR Lyrae stars via the Fourier decomposition and \textit{ad-hoc} calibrations. We also aim to employ the $Gaia$-eDR3 proper motions to distinguish cluster members from field stars in our images of the cluster and to produce a cleaner Colour-Magnitude diagram (CMD) that should enable some observational and theoretical considerations regarding the structure and stellar distribution on the Horizontal Branch (HB). We will also discuss the position of M56 relative to the so-called \textit{Oosterhoff Gap} and its relation with its probable extragalactic origin. Alternative distance determination methods involving the variable stars will be discussed and a theoretical approach of the representation of the HB blue tail as a consequence of mass loss in the RGB will close our work.

\section{Observations and Data Reductions}
\label{observations}

The data used for the present paper were obtained with the 0.84-m Ritchey-Chr\'etien telescope of the Observatorio Astron\'omico Nacional San Pedro M\'artir on two epochs performed approximately one and a half months apart, from August 4-5 2019 for the first, and from September 27-30 2019 for the second. The first set of images consisted of 255 in the $V$ filter and 659 in the $I$ filter while the second set consisted of 258 images in both $V$ and $I$ filters. The estimated average seeing of the first set of images was 1.4"  while for the second set was 1.8". The exposure times ranged between 7s-60s for the first set in both filters while for the second set it ranged between 40s-60s also in both filters. Alongside with the Mexman filter wheel, the detector used was a 2048 x 2048 pixel ESOPO CCD (e2v CCD42-90) with 1.7 e$^{-}$/ADU gain, a readout noise of 3.8 e$^{-}$, a scale of 0.444 arcsec/pix corresponding to a field of view (FoV) of $\approx$ 7.4 arcmin$^{2}$.

\subsection{Difference imaging analysis}

We have employed the software Difference Imaging Analysis (DIA) with its pipeline implementation DanDIA (\citealt{Bramich2008}; \citealt{Bramich2013}) to obtain high-precision photometry of all the point sources in the FoV of our CCD. This allowed us to construct an instrumental light curve for each star. For a detailed explanation on the use of this technique, the reader is referered to the work by \citet{Bramich2011}. 




\subsection{Transformation to the standard system}
Once we obtained the instrumental light curves, we transformed them into the $\emph{VI}$ Johnson-Kron-Cousin standard photometric system \citep{Landolt1992}.
We did this by using 120 standard stars in our FoV identified in the catalog of Photometric Standard Fields \citep{Stetson2000}. The transformation equations \ref{grupoA_V} and \ref{grupoA_I}, correspond to the August 2019 data set, while the transformation equations \ref{grupoB_V} and \ref{grupoB_I} correspond to the September 2019 data set. In all cases there is a mild colour dependence in the transformation from the instrumental to the standard system, particularly for the September data, which we shall use as reference for the zero points of our light curves.

\begin{equation}
    V-v = 0.513(\pm0.017)(v-i)-3.276(\pm0.022)
    \label{grupoA_V}
\end{equation}

\begin{equation}
    I-i=-0.274(\pm0.034)(v-i)-2.056(\pm(0.044))
    \label{grupoA_I}
\end{equation}

\begin{equation}
    V-v = -0.047(\pm0.005)(v-i)-2.585(\pm0.006)
    \label{grupoB_V}
\end{equation}

\begin{equation}
    I-i = 0.051(\pm0.009)(v-i)-2.492(\pm0.012)
    \label{grupoB_I}
\end{equation}




\section{Star membership using $Gaia$-eDR3}
\label{gaia}

\begin{figure*}
\begin{center}
\includegraphics[width=17.0cm, 
height=7.2cm]{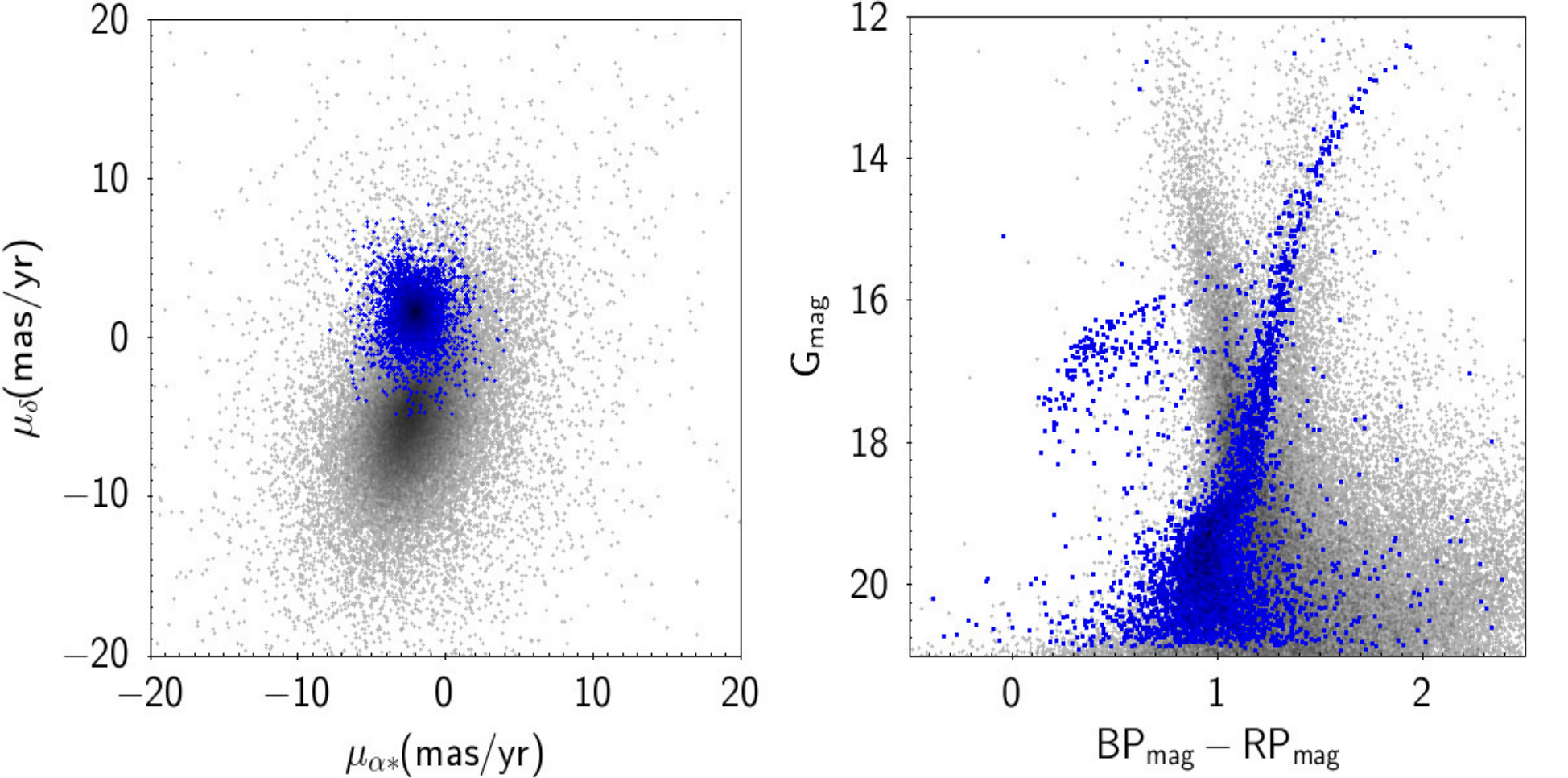}
\caption{The left side panel displays the Vector-point diagram (VPD) of the proper motions of the stars measured by $Gaia$-eDR3 in M56. Stars  within a 25 arcmin radius were used to perform a membership analysis. Blue dots correspond to stars with a high probability of being cluster members while the grey dots correspond to field stars. The panel to the right shows the corresponding CMD where the blue symbols for member stars highlight the characteristic shape of the globular cluster. For details, see Section $\S$ \ref{gaia}.} 
\label{CMD_GAIA}
\end{center}
\end{figure*}

Before studying the CMD of M56, it is convenient to be able to separate field stars from the likely cluster members. We have approached this goal by employing the high-quality astrometric data and proper motions in the $Gaia$-eDR3 data base \citep{Gaia_edr32021} and the method of \citet{Bustos2019}, which is based on
the Balanced Iterative Reducing and Clustering using Hierarchies (BIRCH) algorithm developed by \citet{Zhang1996}. This algorithm detects groups of stars in a 4D space of physical parameters - projection of celestial coordinates ($X_{t}$, $Y_{t}$) and proper motions ($\mu_{\alpha}$, $\mu_{\delta}$). Fig. \ref{CMD_GAIA} illustrates the resulting vector-point diagram (VPD) showing the motion of the cluster relative to the field stars. Also the resulting $Gaia$-CMD of member stars is shown, which is consistent with that of a globular cluster, giving support to the cluster members selection. Unlike more traditional methods, which are based on the fitting of a probability density function and a cut-off at a certain minimum probability to consider stars as reliable cluster members, our method is based on a clustering algorithm as a first stage and a detailed analysis of the residual overdensity as a second stage; member stars extracted in the first stage are labeled M1, and those extracted in the second stage are labeled M2. The analysis was carried out for a 25 arcmin radius field centered in the cluster. We considered 76 235 stars from which 6552 were found to be likely members, out of these only 4695 were in the FoV of our images. We were able to measure and produce light curves for 3632 member stars. 

\begin{figure}
\begin{center}
\includegraphics[width=9.0cm,
height=6.0cm]{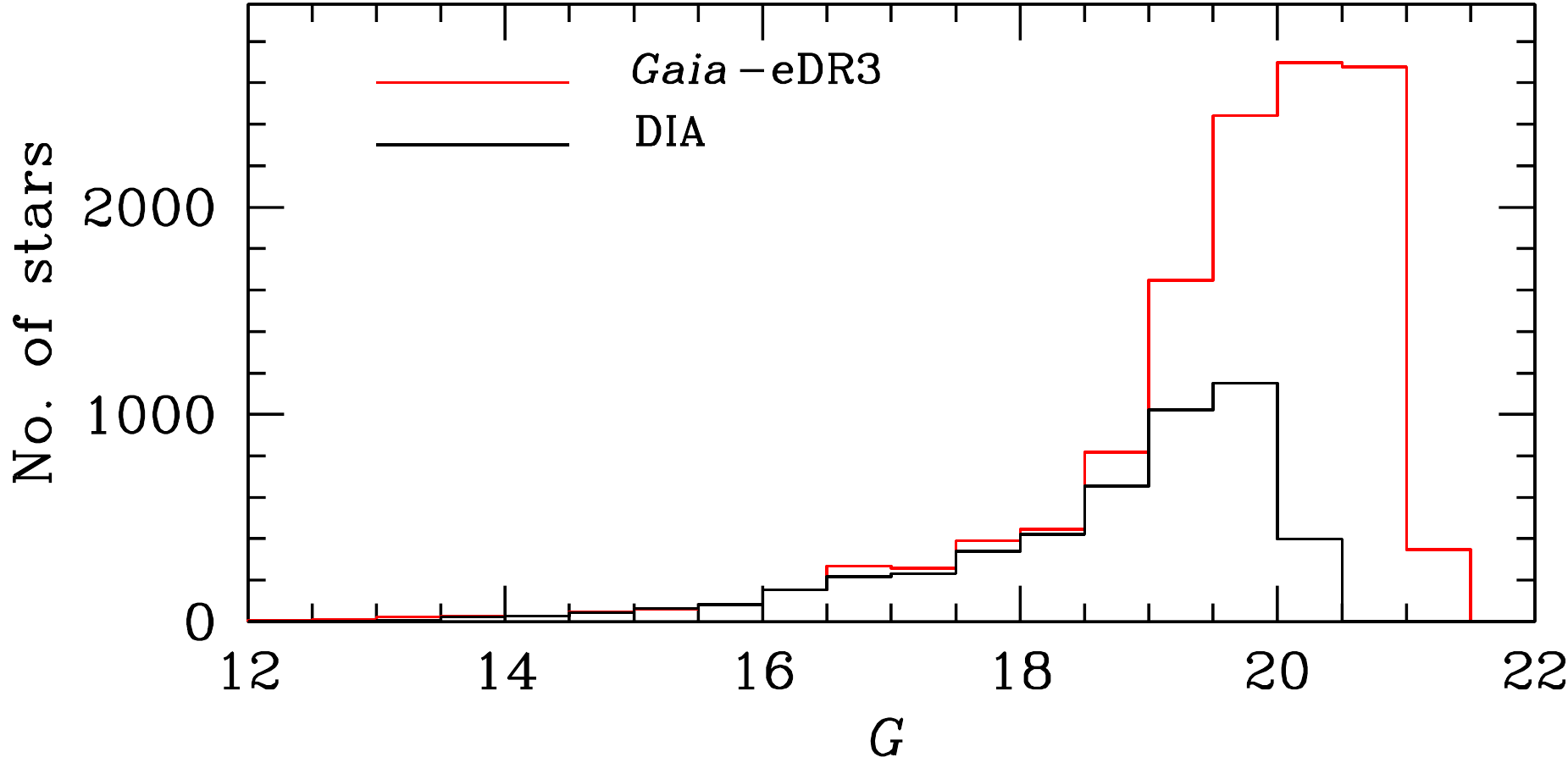}
\caption{Red histogram represents the number of stars in our FoV per $G$ magnitude interval as measured by $Gaia$-eDR3. Black histogram is the magnitude distribution, converted from $V$ to $G$, of the stars measured in this work through our DIA approach. See the discussion on completeness in Section $\S$ \ref{gaia}.} 
\label{completeness}
\end{center}
\end{figure}

From the distribution of field stars in the phase space we estimated the number that is expected to be located in the same region of the sky and of the VPD as the extracted members, therefore they could have been erroneously labelled as members. Within the M1 stars the resulting expected contamination is 77 (1.6$\%$) and within the M2 stars it is 167 (9.0$\%$); therefore, for a given extracted star its probability of being a cluster member is 98$\%$ if it is labeled M1, or 91$\%$ if it is labeled M2.

To evaluate the completeness of our DIA  photometry we compared the number of stars we detected and measured in the FoV of our images, with the ones measured by $Gaia$-eDR3. Of the 76 235 stars considered for the membership analysis, only 12 433 are in our FoV, and we were able to measure 4845. In Fig. \ref{completeness} we confront these two samples as a function of magnitude. We can see that our DIA photometry is complete down to $G$=16.5 mag. which comfortably includes the horizontal portion of the HB, and, it is fairly complete down to $G$=18.5 mag which encompasses the vertical portion of the HB in the CMD.

\section{The variable stars in M56}
\label{var_star}

\begin{figure*}
\begin{center}
\includegraphics[width=\textwidth, height=12cm]{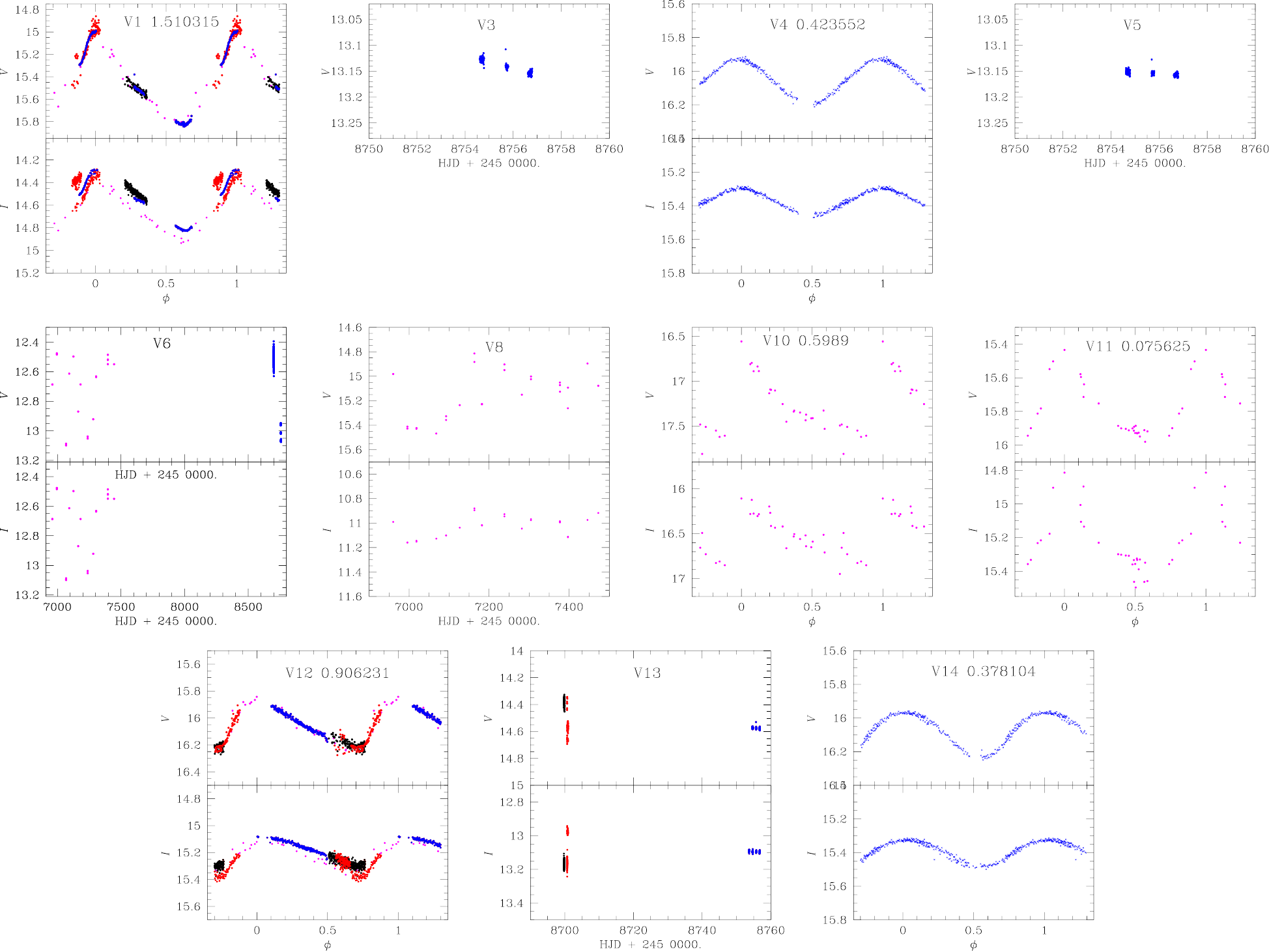}
\caption{Light curves of all known variables in M56 previous to the present work. The colour code is: lilac for data from $Gaia$-DR2 photometry transformed into $V$, black and red correspond to two nights from August 2019, and blue to data from September 2019. Irregular or probably non-periodic giants are plotted as function of HJD. The periods are expressed in fraction of a day.}
\label{knownvar}
\end{center}
\end{figure*}

\begin{figure*}
\begin{center}
\includegraphics[width=\textwidth, height=4cm]{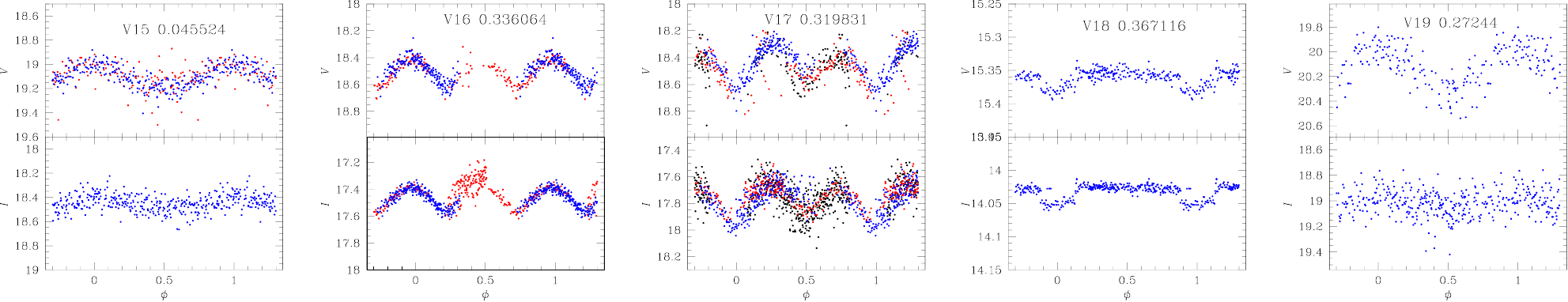}
\caption{Light curves of the variables newly found in this work: 1 SX Phe (V15), 3 EB (V16-V18) and one RRc (V19). The colour code is as in Fig. \ref{knownvar}. The periods are expressed in fraction of a day.}
\label{NEWVAR}
\end{center}
\end{figure*}

\begin{table*}
\scriptsize
\begin{center}
\caption{Data of variable stars in M56 in the FoV of our images.}
\label{variables}

\begin{tabular}{llccccccccc}
\hline
Star ID & Type & $\small<V\small>$ & $\small<I\small>$   & $A_V$  & $A_I$ & $P$    &  $\rm HJD_{\rm max}$ & $\alpha$ (J2000.0)  & $\delta$  (J2000.0)   &  Membership  \\
        &      & (mag) & (mag)   & (mag)  & (mag) & (days) &  + 2450000 &                     &  &status       \\
  
 \hline
V1              &CW     &15.47       &14.60       &0.89  &0.54  &1.510315 &8756.7828 &19:16:39.33 &+30:12:16.6 & M1  \\
V2              &CST?   &--          &--          &--    &--    &--       &--        &19:16:37.34 &+30:11:35.2 & M1 \\
V3              &SR     &13.14$^{4}$ &--          &--    &--    & --      & --       &19:16:37.82 &+30:12:33.9 & M1 \\
V4              &RRc    &16.06       &15.38       &0.28  &0.18  &0.423552 &8700.8311 &19:16:27.45 &+30:08:21.1 &  M1 \\
V5              &SR     &13.15$^{4}$ &--          &--    &      & --      &--        &19:16:36.59 &+30:08:47.3 &  M1 \\
V6              &RVB    &12.75$^{4}$ &11.64$^{4}$ &0.68  &0.62  &--       & --       &19:16:35.77 &+30:11:38.9 &  M1 \\
V7$^{1,2,6}$    &Lb     &--          &--          &--    &--    &--       &--        &19:16:58.74 &+30:07:31.0 &  FS \\
V8$^{1,2}$      &Lb     &15.15$^{5}$ &11.03$^{5}$ &0.66  &0.28  &--       &--        &19:16:28.70 &+30:05:24.2 &  FS \\
V9$^{1,2,6}$    &Lb     &--          &--          &--    &--    &--       &--        &19:16:50.11 &+30:19:58.8 &  FS \\
V10$^{1,2}$     &RRab   &17.26$^{5}$ &16.53$^{5}$ &1.08  &0.73  &0.598900 &7088.5956 &19:16:02.72 &+30:12:25.6 &  FS \\
V11$^{1,2}$     &SX Phe &15.77$^{5}$ &15.20$^{5}$ &0.53  &0.66  &0.075625 &7281.6951 &19:16:03.66 &+30:15:40.5 &  FS \\
V12             &RRab   &16.04       &15.22       &0.42  &0.30  &0.906231 &8753.6200 &19:16:17.25 &+30:09:23.7 &  M1 \\
V13             &SR     &14.58$^{4}$ &13.10$^{4}$ & --   &--     &--       & --      &19:16:38.74 &+30:10:59.0 &  M1 \\
V14             &RRc    &16.09       &15.40       &0.28  &0.17  &0.378104 &8699.8962 &19:16:29.84 &+30:12:27.4 &  M1 \\
V15$^{3}$       &SX Phe &19.11       &18.45       &0.25  &0.23  &0.045524 &8756.7720 &19:16:41.55 &+30:12:08.7 &  M1 \\
V16$^{1,3}$       &EB     &18.50       &17.45       &0.25  &0.20  &0.336064 &8756.7021 &19:16:34.09 &+30:09:09.5 &  FS\\
V17$^{3}$       &EB     &18.47       &17.77       &0.43  &0.39  &0.319831 &8756.6109 &19:16:34.11 &+30:10:25.4 &  M2 \\
V18$^{3}$       &EB     &15.35       &14.03       &--    &--    &0.367116 &8756.7379 &19:16:38.70 &+30:11:09.9 &  M1 \\
V19$^{1,3}$     &RRc    &20.13       &19.02       &0.47  &0.19  &0.273647 &8756.7900 &19:16:30.19 &+30:12:34.5 &  FS\\
\hline
\end{tabular}
\raggedright
\center{
1. Field Star. 2. Out of our FoV.
3. New variable. 4. Magnitude weighted mean. 5. Transformed into $VI$ Johnson standard system from $Gaia$-DR2 photometry. 6. Not measured by $Gaia$-DR2.}

\end{center}
\end{table*}

\subsection{Light curves of known variables}
\label{LC_FV}

Of the 14 known variables in M56, listed in Table \ref{variables}, only V1-V6 and V12-V14 are within the FoV of our images. Variables V7, V8, V9, V10 and V11 are outside the FoV of our images and have been identified as being field variables in the CVSGC and have also been confirmed as such by the membership analysis performed in Section $\S$ \ref{gaia}. For the sake of completeness, we tried to recover their light curves from the $Gaia$-DR2 photometry \citep{Gaia2018} but we were only able to do it for V8, V10 and V11. The variables V7 and V9 do not possess the variability flag associated by $Gaia$ and therefore were not measured. For V8, V10 and V11 we proceeded to transform their $Gaia$ $G$, $G_{BP}$, and $G_{RP}$ bands into the \emph{VI} Johnson standard system to make them consistent with our photometry. For this purpose, we used the relations provided by J.M. Carrasco and which can be found in the $Gaia$-DR2 documentation \footnote{\url{https://gea.esac.esa.int/archive/documentation/GDR2/Data_processing/chap_cu5pho/sec_cu5pho_calibr/ssec_cu5pho_PhotTransf.html}} (2018: Gaia team). The light curves of the known variables and the ones recovered with $Gaia$ are displayed in Fig. \ref{knownvar}.

We remark that the red giants V3, V5 and V6 were saturated in our $I$ images and as such we have no $I$-band photometry.

\subsection{The search for new variables}
The small number of reported variables present in M56 prompted a rigorous search in order to provide a more complete sample for our analysis. We approached this task by using the string-length method (\citealt{Burke1970}, \citealt{Dworetsky1983}). We phased each light curve in our data with periods between 0.02 d and 1.7 d in steps of $10^{-6}$ d. In each case, the length of the line joining consecutive points, called the string-length, represented by the parameter $S_{Q}$ was calculated. The best phasing occurs when $S_{Q}$ is minimum, and corresponds to the best period our data can provide. This method led us to the discovery of five new \textit{bona fide} variables: 1 SX Phe, 3 EB and 1 RRc and were named V15-V19. According to the method described in Section $\S$ \ref{gaia}, V16 (EB) and V19 (RRc) are field variables. Their light curves are displayed in Fig. \ref{NEWVAR}. The ones deemed with a high probability of being cluster members, (coded M1 and M2 in Table \ref{variables}), are in the expected region of the CMD (Fig. \ref{CMD_m56}), in good agreement with their variability type. All variables listed in Table \ref{variables} that are contained in our FoV are identified in the cluster chart of Fig. \ref{idchart}.

\subsection{Comments on the reddening of M56 }

The canonical reddening generally quoted for M56 is $E(B-V)=0.26$ \citep{Harris1996}. In an attempt to confirm this value we have considered V12, the only RRab present in M56. We have made used of the fact that RRab stars have a nearly constant intrinsic colour $(B-V)_0$, between phases 0.5 and 0.8 \citep{Sturch1966}. The intrinsic value $(V-I)_0$ in this range of phases was calibrated  by \citet{Guldenschuh2005} who found a value of  $\overline{(V-I)_{o,min}}$ = 0.58 $\pm$ 0.02 mag. Hence, by measuring the average $\overline{(V-I)_{o,0.5-0.8}}$, one can estimate $E(V-I) =\overline{(V-I)_{o,min}}$ -  $(V-I)_{o,0.5-0.8}$ for a given RRab star with a properly covered colour curve at these phases. Then, we calculated $E(B-V)$ via the ratio $E(V-I)/E(B-V) = 1.259$.

While our light curve of V12 is not covered near its maximum (see Fig. \ref{knownvar}), the 0.5 - 0.8 phase range is very well covered. We calculated $E(B-V)=0.26\pm0.02$. The dust map calibrations of \citet{Schlafly2011} and \citet{Schlegel1998} give values of $0.216 \pm0.009$ and $0.252 \pm0.011$, respectively. A large range of reddening values can be found in the literature, from 0.18 \citep{Ivanov2000} to  0.32 \citep{Hatzidimitriou2004}).

For the rest of the paper we shall adopt the value of $E(B-V)=0.26$.

\section{The Fourier decomposition approach to the physical parameters of RR Lyrae stars}\label{fourier}

The use of Fourier light curve decomposition has been a well tested approach towards the determination of key physical parameters of RR Lyrae stars. The amplitudes and displacements, $A_{k}$ and $\phi_{k}$, and the period $P$ of a series of harmonics of the form:

\begin{equation}
    m(t) = A_{0} + \sum_{k=1}^{N}A_{k}cos(\frac{2\pi}{P}k(t-E_{0}) + \phi_{k}),
	\label{quadratic}
\end{equation}

\noindent
are intimately correlated with stellar values of 
[Fe/H], distance, and other relevant physical parameters. In eq. \ref{quadratic}, $m(t)$ is the magnitude of the star at time $t$, $P$ is the period of pulsation, and $E_0$ is the epoch. To calculate $A_k$ and  $\phi_k$ of each harmonic, we made used of a least-squares fit approach. The Fourier parameters, are defined as $\phi_{ij} = j\phi_{i} - i\phi_{j}$, and $R_{ij} = A_{i}/A_{j}$. For the RR Lyrae stars, their Fourier coefficients are listed in Table \ref{fourier_coeffs}.

\noindent
These parameters, inserted in \textit{ad-hoc} calibrations for RRab and RRc stars, can lead to individual values of the physical parameters reported in Table \ref{fisicos}. The specific calibrations and their zero points are given and discussed to a great detail in several of our previous papers (e.g. \citet{Are2017} and \citet{Deras2019}) and shall not be repeated here.

M56 is a particularly weak case for Fourier decomposition since it contains a very limited sample of member RR Lyrae: one RRab (V12) and two RRc (V4 and V14). The value of [Fe/H]$_{\rm ZW}$, i.e., the metallicity in the scale of \citet{Zinn1984}, is obtained from the calibration of \citet{Morgan2007} for RRc stars and \citet{Jurcsik1996} for the RRab star. A value in the high resolution spectroscopic scale [Fe/H]$_{\rm spec}$ can be obtained from the calibrations of \citet{Nemec2013}. In all cases, the key Fourier parameters are $\phi_{31}$ and $P$. For V4, the value $\phi_{31}$=5.730 (Table \ref{fourier_coeffs}) is far outside the range of the RRc stars that define the calibration, or in fact it is anomalously large among RRc stars in general (e.g. see Fig. 2 of \citet{Clement1992}). This fact is most likely due to its peculiar light curve shape, with maxima more acute than in typical RRc stars (compare with the light curve of V14). For the RRab star V12 we also face some inconvenience: the value of its compatibility parameter, as defined by \citet{Jurcsik1996} and \citet{Kovacs1998} is $D_m$=4.3. These authors recommend values smaller than 3.0 to warrant the consistency of a given light curve with the morphology of those of the calibrators. Hence V12 seems also a bit off the limits.

As a result we have a metallicity estimation via the Fourier approach for one RRc star (V14); [Fe/H]$_{\rm Spec}$=2.03$\pm$0.14. This value is, however, consistent with the standard value reported by \citet{Harris1996} of $-1.98$.

\begin{table*}
\footnotesize
\caption{Fourier coefficients for the RR Lyrae stars in the FoV of our images.}
\centering                   
\begin{tabular}{lllllllllr}
\hline
Variable ID     & $A_{0}$    & $A_{1}$   & $A_{2}$   & $A_{3}$   & $A_{4}$   &
$\phi_{21}$ & $\phi_{31}$ & $\phi_{41}$ 
&  $D_{\mbox{\scriptsize m}}$ \\
     & ($V$ mag)  & ($V$ mag)  &  ($V$ mag) & ($V$ mag)& ($V$ mag) & & & & \\
\hline
       &     &   &   & RRab  && &  &  &       \\
\hline
V12 &16.044(2) &0.159(3) &0.053(3) &0.020(3) &0.010(2) &4.451 &9.304 &8.388 & 4.3 \\
\hline
        &           &      &           &     RRc      & &&&&\\
\hline
V4 &16.062(1) &0.130(1) &0.002(1) &0.003(1) &0.002(1) &4.442 &5.730 &4.380 & -- \\
V14 &16.092(1) &0.134(1) &0.013(1) & 0.004(1)&0.003(1) &5.583 &3.516 &2.939 &--  \\
\hline	
\hline
\end{tabular}
\label{fourier_coeffs}
\end{table*}

\begin{table*}
\footnotesize
\begin{center}
\caption{Physical parameters obtained from the Fourier fit for the \RRab ~and \RRc ~stars.}
\label{fisicos}
\hspace{0.01cm}
 \begin{tabular}{lllllllll}
\hline
Star&[Fe/H]$_{\rm ZW}$&[Fe/H]$_{\rm Spec}$ & $M_V$ & log~$T_{\rm eff}$  & log$(L/{L_{\odot}})$ & $M/{M_{\odot}}$&$R/{R_{\odot}}$&d(kpc)\\
\hline 
 &  &  & RRab star &  & & &\\
\hline
V12  & -1.76(5) & -1.74(6) & 0.323(4) & 3.808(1) & 1.771(2) & 0.467(3)  &  6.25(1)  & 9.62(2)\\
\hline
 &  &  & RRc stars &  & & &\\

\hline
V4  & -- & -- & 0.52(3) &3.848(2)  &1.693(11) & 0.46(2)&   4.74(6)& 8.86(11)\\

V14  & --1.96(9) &--2.03(14) &  0.51(1) &3.846(2) &1.697(2) & 0.57(1) &4.81(1) & 9.03(3) \\

\hline
Weighted Mean &--1.96 &-2.03 &0.51 &3.847 &1.696 &0.54 & 4.81&9.02\\
\hline
\end{tabular}
\end{center}
\raggedright
\center{
The numbers in
parentheses indicate the uncertainty on the last decimal place.\\
 Also listed is the deviation parameter $D_{\mbox{\scriptsize m}}$ for the \RRab ~   star. \\
See Section $\S$ \ref{fourier} for a detailed discussion. 
}
\end{table*}

\section{Comments on the distance to M56}
\label{distance}

The distance to M56 has been estimated by several independent approaches, namely, the absolute magnitude estimation of the RR Lyrae stars via the Fourier decomposition of their light curves with the results V12 (9.6 kpc), V4 (8.9 kpc), V14(9.0 kpc) (see also Table \ref{fisicos}); the positioning of the isochrones and ZAHB on the intrinsic CMD (10 kpc); the P-L in the $I$ filter for the RR Lyrae stars (9.0 kpc) \citep{Catelan2004} and the application of three independent and well known P-L relations of SX Phe stars (\citealt{Poretti2008}, \citealt{Arellano2011}, and \citealt{CohenSara2012}). The calculation for the SX Phe stars V11 and V15 (new variable) deserves a further discussion. Assuming that these two stars are pulsating in the fundamental mode, these calibrations lead to the distances of 3.46 $\pm$ 0.14 kpc and 11.5 $\pm$ 0.4 kpc, respectively. First or second overtone pulsation assumptions lead to larger distances.
It is clear, then, that V11 is definitively not a cluster member, in agreement with the determination from the $Gaia$-eDR3 proper motions described in Section $\S$ \ref{gaia}. V15 seems to lie about 1-2 kpc behind the cluster and may or may not belong to the cluster, hence we did not include it in the calculation of the average distance shown in Table \ref{tabdistance}.

\begin{table}
\begin{center}
\caption{Distance to M56 from different methods} 
\label{tabdistance}
\begin{tabular}{lll}
\hline
Method&$d$ (kpc)&N\\
\hline
RRab Fourier decomposition & 9.62 &1\\
RRc Fourier decomposition  & 8.95$\pm$0.09&2\\
ZAHB placement on the CMD  & 10.0$\pm$0.2 \\
RRLs P-L ($I$) & 9.0$\pm$0.2&3\\
SX Phe P-L & 11.5$^{*}$&1\\
\hline
Average&9.39$\pm$0.44& \\
\hline
\end{tabular}

\center{\quad $*$ Not included in average}
\end{center}
\end{table}

\section{the cmd of M56 and its HB}
\label{cmd}
\begin{figure*}
\begin{center}
\includegraphics[width=\textwidth, height=8cm]{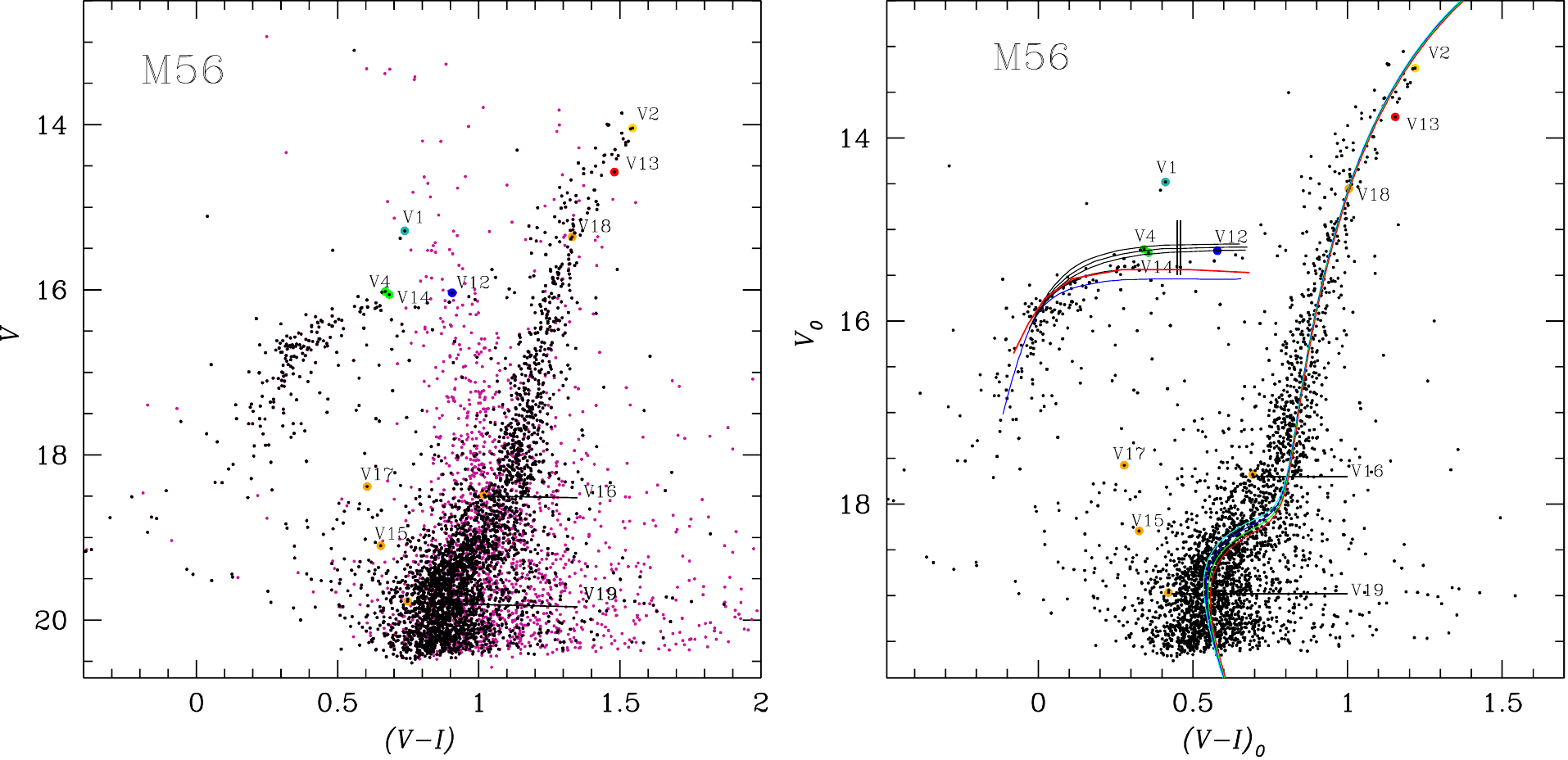}
\caption{The Colour-Magnitude Diagram of M56. The left panel shows all the stars measured in the FoV of our images. Non-member field stars identified by the method described in Section $\S$ \ref{gaia} are shown in purple colour, while the likely members are represented by black dots. The colour code for the variable stars is as follows: in teal the CW, in green the RRc, in blue the RRab, in orange the newly discovered variables, in red the SR, and in yellow a suspected but still unconfirmed variable. The right panel is the intrinsic CMD for $E(B-V)=0.26$, where isochrones for ages 12.0 (cyan), 12.4 (blue), 13.0 (green) and 13.5 (red) Gyrs, [Fe/H]=--2.0, Y=0.25 and [$\alpha$/Fe]=+0.4, built from the models of \citet{Vandenberg2014}, have been added by shifting them to a distance of 10 kpc to reproduce the observations. The red ZAHB was calculated from  the  Eggleton  code  (\citealt{Pols1997}; \citealt{Pols1998};  \citealt{KPS1997}). The blue ZAHB is from the models of \citet{Vandenberg2014}. The  thin  black  lines  are  evolutionary  tracks  for a helium core mass of 0.50 M$_{\odot}$ and total masses of  0.64, 0.65  and  0.66 M$_{\odot}$ also from  the  Eggleton  code. The empirical border between first overtone and fundamental mode, i.e., the red edge of the first overtone instability strip (FORE), is shown as two vertical black lines.  (See Section $\S$ \ref{cmd} for a full discussion).}
\label{CMD_m56}
\end{center}
\end{figure*}

\begin{figure*}
\begin{center}
\includegraphics[width=14cm, height=14cm]{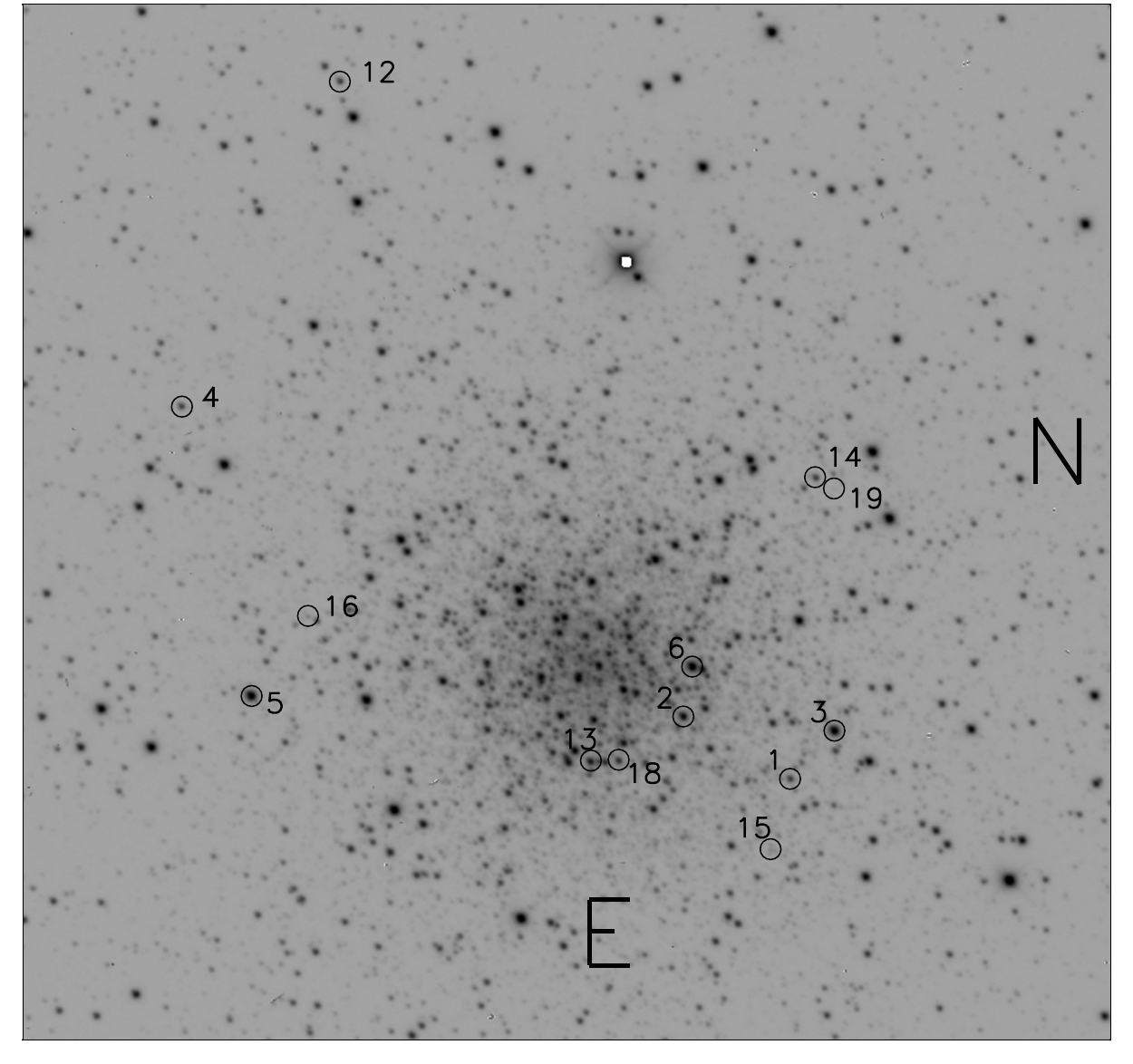}
\caption{Identification chart of variables in our FoV of M56. The field is 5.92$\times$5.92 arcmin$^{2}$. Expansion of the digital version is recommended for clearness.}
\label{idchart}
\end{center}
\end{figure*}

\begin{figure}
\begin{center}
\includegraphics[width=8.2cm, height=8.2cm]{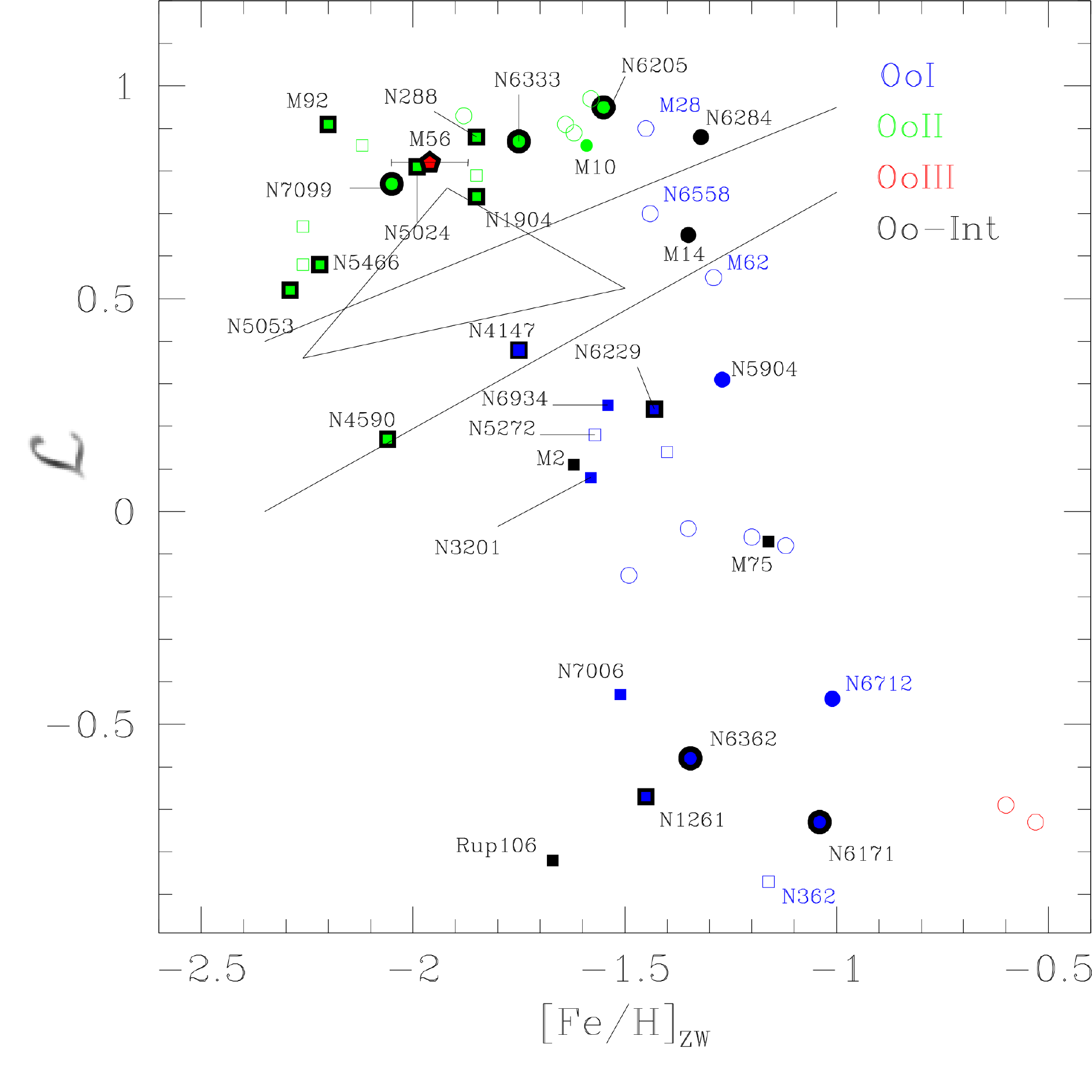}
\caption{The HB structure parameter $\mathcal L$ as a function of [Fe/H]$_{\rm ZW}$. The two lines define theoretically the so-called Oosterhoff Gap \protect\citep{Bono1994}, a region in the $\mathcal L$ - [Fe/H]$_{\rm ZW}$ plane that is devoid of \textit{bona fide} Galactic globular clusters. The triangle represents a region found by \protect\citet{Catelan2009} that is populated by globulars identified in systems outside of the Galaxy. M56 is coloured in red and its position was determined by the estimated values of the metallicity and $\mathcal L$ in this work (0.82, -1.96).
Circles and squares are used for inner and outer halo clusters, respectively. Symbols with a black rim represent globular clusters where the fundamental and first modes of pulsations are well segregated on the HB, while in those without the modes of pulsation are mixed in the either/or region. Open symbols correspond to clusters yet to be studied by our group and their current positions are subject to change. For a detailed discussion, see Section $\S$ \ref{cmd}.}

\label{L_parameter}
\end{center}
\end{figure}

From the PSF-fitting photometric analysis of our images in Section $\S$ \ref{observations}, we were able to recover 4845 stars, 3630 ($\sim$ 75$\%$) of which have a high probability to be cluster members. This allowed us to create a clean CMD by removing the field stars (Fig. \ref{CMD_m56}). We used the models from \citet{Vandenberg2014} to generate a ZAHB  and four isochrones of ages ranging from 12.0 - 13.5 Gyrs with a metallicity of [Fe/H]$_{\rm ZW}$ = -2.0, Y=0.25 and [$\alpha$/Fe]=0.4. Also, using the Eggleton code (\citealt{Pols1997}; \citealt{Pols1998};  \citealt{KPS1997}) we generated another ZAHB for a core mass of 0.50 M$_{\odot}$, and three evolutionary tracks corresponding to total masses of 0.64, 0.65, and 0.66 M$_{\odot}$. The isochrones and the ZAHBs were shifted to a distance of 10 kpc using a reddening $E(B-V)$ = 0.26. The small discrepancy between the ZAHB from the models of VandenBerg and Eggleton is due to the assumed core mass of the progenitor star.\\
Although the sample of RR Lyrae stars is small, the fact that the RRc and RRab stars are on the first overtone and fundamental sides of the First Overtone Blue Edge (FOBE), is consistent with the mode distribution found in other OoII type clusters, e.g. \citet{Yepez2020}. A useful quantity to help describe the morphology of the HB is the Lee parameter \citep{Lee1990} defined as $\mathcal L$ = $(B-R) / (B+V+R)$, where $B$ is the number of stars on the blue side of the Instability Strip (IS) or th First Overtone Blue Edge (FOBE), $R$ is the number of stars on the red side of the IS or Fundamental Mode Red Edge (FMRE) and $V$ is the number of variable stars within the IS. In the case of M56, this yields an $\mathcal L$ = 0.82. 
The Fig. \ref{L_parameter} shows a plot of $\mathcal L$ vs [Fe/H]$_{\rm ZW}$ where globular clusters previously studied by our group following the same approach as the one used in this work, are plotted. The triangle in Fig. \ref{L_parameter} is defined by \citet{Catelan2009} as a region that is preferentially populated by clusters thought to be of extragalactic origin. M56 falls well above this triangular region. The HB of M56 is predominately blue, which is usually seen in most metal-poor Oosterhoff II (OoII) type clusters. While it is not possible to discuss the Oo-type of M56 given the scarce number of RR Lyrae stars, the extremely large period of the only RRab star V12, makes the case most peculiar even among Oo-II type clusters. Its position on the $\mathcal L$ vs [Fe/H]$_{\rm ZW}$ plane however makes it difficult to support its possible extragalactic origin on this grounds.

\subsection{Post-evolutionary ZAHB models}
\label{sec:zahb_models}
\begin{figure}
\begin{center}
\includegraphics[width=8cm, height=8cm]{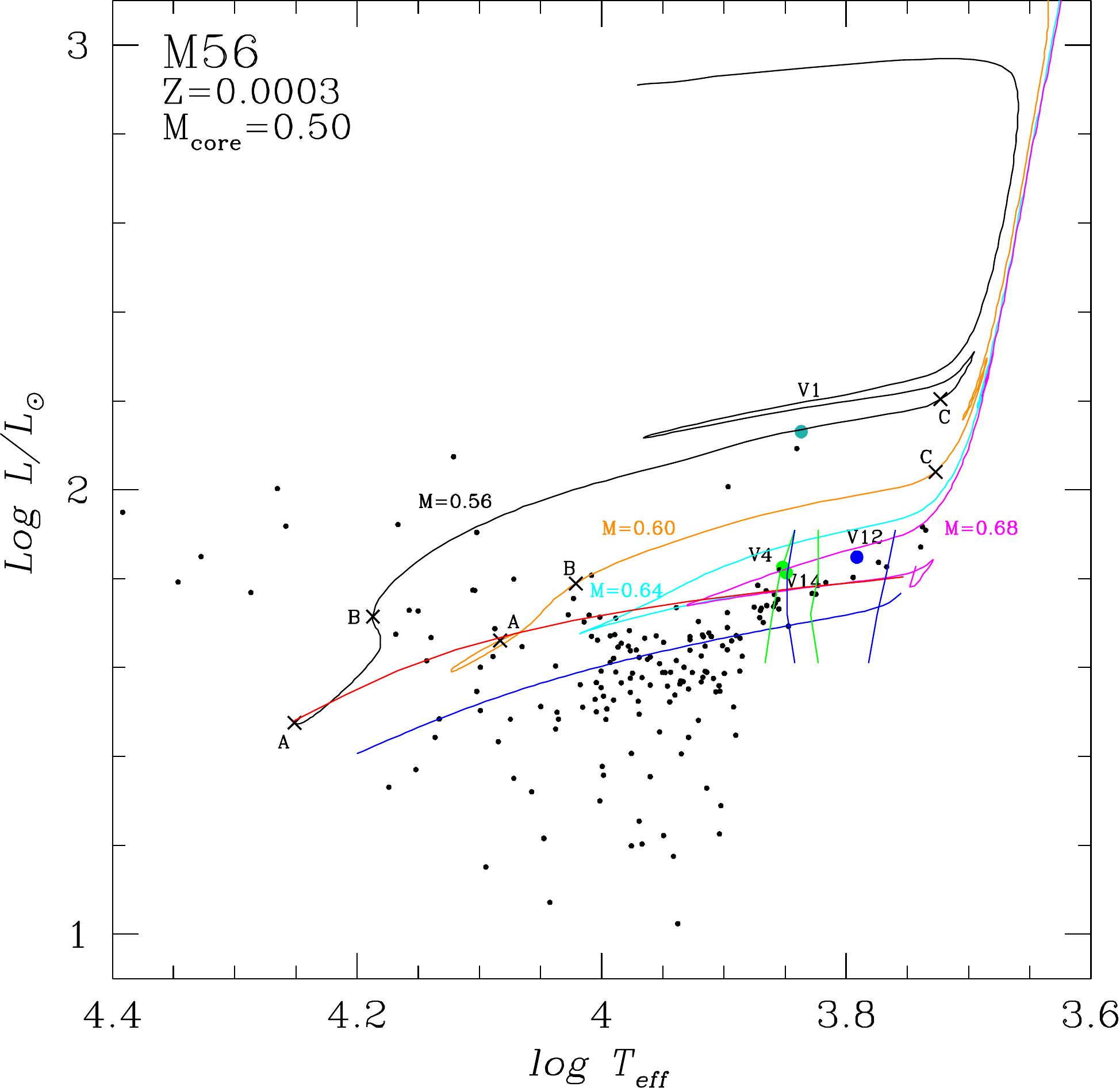}
\caption{HRD of the Horizontal Branch of M56. These models were created using a Helium core mass of 0.50 M$_{\odot}$. 
Four tracks for total masses of 0.56 (black), 0.60 (orange), 0.64 (cyan) and 0.68 M$_{\odot}$ (lilac) are shown. The blue ZAHB is from the models of \citet{Vandenberg2014}. Theoretical border lines of the IS are from \citet{Bono1994} for the fundamental mode (blue) and the first overtone (green). The colour code of the variable stars is as in Fig. 4. The fiducial marks at A ($\approx$ 2.5 Myrs), B (leaving the HB) and C (at the base of the AGB) along the tracks help to visualise the evolutionary times involved; on the 0.56 M$_{\odot}$ track, the time from A to C is about 100.9 Myrs, thus the post-ZAHB age of V1 is about 100.8 Myrs. The crossing time from B to C takes roughly 2 Myrs.}
\label{HRD_m56}
\end{center}
\end{figure}

In order to offer some input on the stellar mass distribution on the HB, as well as on the mass structure (core and envelope masses) for the stars of the HB and their ulterior evolution towards higher luminosities, we have calculated some evolutionary tracks using the Eggleton code (\citealt{Pols1997}; \citealt{Pols1998}; \citealt{KPS1997}) with a modified Reimers law for the mass loss with $\eta = 0.8 \times 10^{-13}$ \citep{Schroeder2005} as a consequence of the He-flash episodes during the RGB stages. A detailed description of this procedure has been given by \citet{Arellano2020}.
To compare our theoretical predictions with the observed stellar distributions, we converted the CMD into an HRD, i.e., we converted the plane $(V-I)$-$V$ into log($T_{\rm eff}$)-log$(L/{L_{\odot}})$. To this end, we adopted the $(V-I)_o$-log($T_{\rm eff}$) and $BC$-log($T_{\rm eff}$) calibrations of \citet{Vandenberg2003}. The resulting HRD of the HB region is shown in Fig. \ref{HRD_m56}. It should be noted that the scatter in HB is larger on its blue side, i.e., the stars corresponding to the vertical blue tail in the HB in the CMD. Since they are considerably fainter, some may or may not be truly members of the HB. In comparison, the horizontal portion where the RR Lyrae stars reside is neat and well represented by the theoretical loci. While in fact we have modelled cases for core masses of 0.49, 0.50 and 0.51 M$_\odot$, we display only the case for 0.50 M$_\odot$ as we believe they produced the best representation of the data.

In Fig. \ref{HRD_m56} there are also shown the positions of the RRab V12 (blue dot), RRc V4 and V14 stars (green dots) and the only CW V1 star (teal dot), along with the theoretical borders of the instability strip for the fundamental and first overtone modes \citep{Bono1994}. The red line is the ZAHB for the stars with a He-core of 0.50 M$_\odot$. The blue line corresponds to the ZAHB calculated with the models of \citet{Vandenberg2014} for [Fe/H]=-2.0, Y=0.25 and $\rm [\alpha/Fe]$=+0.4; this ZAHB starts with a core mass of 0.49 M$_\odot$ that decreases towards the blue and for this reason it lies a bit below of the red line. Black, orange, cyan and lilac loci  are tracks for total masses of 0.56, 0.60, 0.64, 0.68 M$_\odot$, respectively, i.e. for envelopes with 0.06, 0.10, 0.14 and 0.18 M$_\odot$. The RR Lyrae stars lie between tracks with envelopes of 0.14-0.18 M$_\odot$. The CW star, V1, seems well represented by a track with a very thin envelope (0.06 M$_\odot$) which has been originated at the bluest part of HB. \\
Considering an age range of 12.0 - 13.5 Gyrs, the probable progenitor star of the stars on the HB had an initial mass  of $\sim$ 0.81 - 0.83 M$_\odot$ at the ZAMS. Such star lost between 0.15 and 0.26 M$_\odot$ during the He flashes in its evolution from the RGB towards the ZAHB.

\section{summary}
\label{summary}

In this work we have performed $VI$ CCD photometry of 4845 sources present in the globular cluster M56 and we summarise our results as follows:
\begin{itemize}
    
    \item The use of a semi-empirical calibration based on the Fourier decomposition of the light curve of a single RRc star (V14) allowed us to determine a metallicity value of [Fe/H]$_{\rm ZW}$ = -1.96 $\pm$ 0.09.   
    
    \item The fact that the intrinsic color $(B-V)_{0}$ of RRab stars at minimum light between the phases 0.5 - 0.8 is constant, allowed us to estimated a value for the reddening using the only RRab present in our data (V12), yielding $E(B-V)$ = 0.26. 
    
    \item We have made use of five different independent methods to determine the distance to M56, namely: from the semi-empirical calibrations derived for RRab and RRc stars, from the placement of a ZAHB on the CMD, from the P-L relation in the $I$ filter for RR Lyrae stars, and the P-L relation for SX Phe stars. For the last method, our membership analysis determined V15 to be a cluster member, and its position on the CMD is consistent with the position of the stars of the same class. Nevertheless, its estimated distance (assuming fundamental mode pulsation) places it approximately 2 kpc farther than the distance obtained through the other four methods ($\big<d\big>$ = 9.39 $\pm$ 0.44 kpc), and was not included in the calculation of the average cluster distance.
    
    \item Our analysis of star membership through proper motions using $Gaia$-eDR3, allowed us to create a clean CMD consistent with isochrones in an age range of 12.0-13.5 Gyrs and a metallicity of [Fe/H]$_{\rm ZW}$ = -2.0.
    
    \item Previous authors have presented evidence that indicates that the origin of M56 is the result of a merger event. The value of the $\mathcal{L}$ parameter, calculated from a CMD clean of field stars, (0.82) and the value of the metallicity of M56 (-1.96), places M56 clearly outside the region of the $\mathcal{L}$ vs [Fe/H]$_{\rm ZW}$ plane preferentially populated by globular clusters of extragalactic origin. These results make it difficult to argue in favour the extragalactic origin of M56.
    
    \item Our post He flash evolutionary models with a He-core of 0.50 M$_\odot$ and envelopes of 0.04 - 0.18 M$_\odot$, cover the complete HB of M56. Models with large total masses ($\sim$ 0.68 M$_\odot$) can explain the evolutionary stage of the RR Lyrae stars. Models with thinner envelopes show longer blue loops that cross the IS above the HB, in the region of the type II Cepheids. Stars in the blue tail with these features could be the origin of BL Her and W Vir stars.

    \item We report five new variables: 1 SX Phe (V15), 3 EB (V16 - V18), and 1 RRc (V19). According to the proper motion analysis carried out in Section $\S$ 3, V16 (EB) and V19 (RRc) are not cluster members. In particular for V19, this result is confirmed by the value of its intensity weighted mean ($V$ $\sim$ 20) and its unusual location at the bottom of the CMD, more than three magnitudes below the HB.
\end{itemize}

\section{ACKNOWLEDGMENTS}
This project was partially supported by DGAPA-UNAM (Mexico) via grants IG100620 and  IN106615-17. We have made extensive use of the SIMBAD, ADS services, and of `Aladin sky atlas' developed at CDS, Strasbourg Observatory, France \citep{Bonnarel2000}, and
TOPCAT \citep{Taylor2005}.

\bibliographystyle{rmaa}
\bibliography{biblioM56}

\begin{thebibliography}
\expandafter\ifx\csname natexlab\endcsname\relax\def\natexlab#1{#1}\fi
\expandafter\ifx\csname href\endcsname\relax
  \def\href#1#2{}\fi
\expandafter\ifx\csname urllinklabel\endcsname\relax
  \def\urllinklabel{[LINK]}\fi
\expandafter\ifx\csname adsurllinklabel\endcsname\relax
  \def\adsurllinklabel{[ADS]}\fi

\bibitem[{{Arellano Ferro} {et~al.}(2017){Arellano Ferro}, {Bramich}, \&
  {Giridhar}}]{Are2017}
{Arellano Ferro}, A., {Bramich}, D.~M., \& {Giridhar}, S. 2017, \rmxaa, 53, 121


\bibitem[{{Arellano Ferro} {et~al.}(2011){Arellano Ferro}, {Figuera Jaimes},
  {Giridhar}, {Bramich}, {Hern{\'a}ndez Santisteban}, \&
  {Kuppuswamy}}]{Arellano2011}
{Arellano Ferro}, A., {Figuera Jaimes}, R., {Giridhar}, S., {Bramich}, D.~M.,
  {Hern{\'a}ndez Santisteban}, J.~V., \& {Kuppuswamy}, K. 2011, \mnras, 416,
  2265


\bibitem[{{Arellano Ferro} {et~al.}(2020){Arellano Ferro}, {Yepez}, {Muneer},
  {Bustos Fierro}, {Schr{\"o}der}, {Giridhar}, \&
  {Calder{\'o}n}}]{Arellano2020}
{Arellano Ferro}, A., {Yepez}, M.~A., {Muneer}, S., {Bustos Fierro}, I.~H.,
  {Schr{\"o}der}, K.~P., {Giridhar}, S., \& {Calder{\'o}n}, J.~H. 2020, \mnras,
  499, 4026


\bibitem[{{Bonnarel} {et~al.}(2000){Bonnarel}, {Fernique}, {Bienaym{\'e}},
  {Egret}, {Genova}, {Louys}, {Ochsenbein}, {Wenger}, \&
  {Bartlett}}]{Bonnarel2000}
{Bonnarel}, F., {Fernique}, P., {Bienaym{\'e}}, O., {Egret}, D., {Genova}, F.,
  {Louys}, M., {Ochsenbein}, F., {Wenger}, M., \& {Bartlett}, J.~G. 2000,
  \aaps, 143, 33


\bibitem[{{Bono} {et~al.}(1994){Bono}, {Caputo}, \& {Stellingwerf}}]{Bono1994}
{Bono}, G., {Caputo}, F., \& {Stellingwerf}, R.~F. 1994, \apj, 423, 294


\bibitem[{{Bramich}(2008)}]{Bramich2008}
{Bramich}, D.~M. 2008, \mnras, 386, L77


\bibitem[{{Bramich} {et~al.}(2011){Bramich}, {Figuera Jaimes}, {Giridhar}, \&
  {Arellano Ferro}}]{Bramich2011}
{Bramich}, D.~M., {Figuera Jaimes}, R., {Giridhar}, S., \& {Arellano Ferro}, A.
  2011, MNRAS, 413, 1275


\bibitem[{{Bramich} {et~al.}(2013){Bramich}, {Horne}, {Albrow}, {Tsapras},
  {Snodgrass}, {Street}, {Hundertmark}, {Kains}, {Arellano Ferro}, {Figuera},
  \& {Giridhar}}]{Bramich2013}
{Bramich}, D.~M., {Horne}, K., {Albrow}, M.~D., {Tsapras}, Y., {Snodgrass}, C.,
  {Street}, R.~A., {Hundertmark}, M., {Kains}, N., {Arellano Ferro}, A.,
  {Figuera}, J.~R., \& {Giridhar}, S. 2013, \mnras, 428, 2275


\bibitem[{{Burke} {et~al.}(1970){Burke}, {Rolland}, \& {Boy}}]{Burke1970}
{Burke}, Edward~W., J., {Rolland}, W.~W., \& {Boy}, W.~R. 1970, Journal of the
  Royal Astronomical Society of Canada, 64, 353


\bibitem[{{Bustos Fierro} \& {Calder{\'o}n}(2019)}]{Bustos2019}
{Bustos Fierro}, I.~H. \& {Calder{\'o}n}, J.~H. 2019, \mnras, 488, 3024


\bibitem[{{Catelan}(2009)}]{Catelan2009}
{Catelan}, M. 2009, \apss, 320, 261


\bibitem[{{Catelan} {et~al.}(2004){Catelan}, {Pritzl}, \&
  {Smith}}]{Catelan2004}
{Catelan}, M., {Pritzl}, B.~J., \& {Smith}, H.~A. 2004, \apjs, 154, 633


\bibitem[{{Clement} {et~al.}(1992){Clement}, {Jankulak}, \&
  {Simon}}]{Clement1992}
{Clement}, C.~M., {Jankulak}, M., \& {Simon}, N.~R. 1992, \apj, 395, 192


\bibitem[{{Clement} {et~al.}(2001){Clement}, {Muzzin}, {Dufton}, {Ponnampalam},
  {Wang}, {Burford}, {Richardson}, {Rosebery}, {Rowe}, \& {Hogg}}]{Clement2001}
{Clement}, C.~M., {Muzzin}, A., {Dufton}, Q., {Ponnampalam}, T., {Wang}, J.,
  {Burford}, J., {Richardson}, A., {Rosebery}, T., {Rowe}, J., \& {Hogg}, H.~S.
  2001, \aj, 122, 2587


\bibitem[{{Cohen} \& {Sarajedini}(2012)}]{CohenSara2012}
{Cohen}, R.~E. \& {Sarajedini}, A. 2012, \mnras, 419, 342


\bibitem[{{Deason} {et~al.}(2013){Deason}, {Belokurov}, {Evans}, \&
  {Johnston}}]{Deason2013}
{Deason}, A.~J., {Belokurov}, V., {Evans}, N.~W., \& {Johnston}, K.~V. 2013,
  \apj, 763, 113


\bibitem[{{Deras} {et~al.}(2019){Deras}, {Ferro}, {L{\'a}zaro}, {Fierro},
  {Calder{\'o}n}, {Muneer}, \& {Giridhar}}]{Deras2019}
{Deras}, D., {Ferro}, A.~A., {L{\'a}zaro}, C., {Fierro}, I.~H.~B.,
  {Calder{\'o}n}, J.~H., {Muneer}, S., \& {Giridhar}, S. 2019, \mnras, 626


\bibitem[{{Dworetsky}(1983)}]{Dworetsky1983}
{Dworetsky}, M.~M. 1983, \mnras, 203, 917


\bibitem[{{Gaia Collaboration}(2021)}]{Gaia_edr32021}
{Gaia Collaboration}. 2021, \aap, 649, A1


\bibitem[{{Gaia Collaboration} {et~al.}(2018){Gaia Collaboration}, {Brown},
  {Vallenari}, {Prusti}, {de Bruijne}, {Babusiaux}, {Bailer-Jones}, {Biermann},
  {Evans}, {Eyer}, \& et~al.}]{Gaia2018}
{Gaia Collaboration}, {Brown}, A.~G.~A., {Vallenari}, A., {Prusti}, T., {de
  Bruijne}, J.~H.~J., {Babusiaux}, C., {Bailer-Jones}, C.~A.~L., {Biermann},
  M., {Evans}, D.~W., {Eyer}, L., \& et~al. 2018, \aap, 616, A1


\bibitem[{{Guldenschuh} {et~al.}(2005){Guldenschuh}, {Layden}, {Wan},
  {Whiting}, {van der Bliek}, {Baca}, {Carlin}, {Freismuth}, {Mora}, {Salyk},
  {Vera}, {Verdugo}, \& {Young}}]{Guldenschuh2005}
{Guldenschuh}, K.~A., {Layden}, A.~C., {Wan}, Y., {Whiting}, A., {van der
  Bliek}, N., {Baca}, P., {Carlin}, J., {Freismuth}, T., {Mora}, M., {Salyk},
  C., {Vera}, S., {Verdugo}, M., \& {Young}, A. 2005, \pasp, 117, 721


\bibitem[{{Harris}(1996)}]{Harris1996}
{Harris}, W.~E. 1996, \aj, 112, 1487


\bibitem[{{Hatzidimitriou} {et~al.}(2004){Hatzidimitriou}, {Antoniou},
  {Papadakis}, {Kaltsa}, {Papadaki}, {Papamastorakis}, \&
  {Croke}}]{Hatzidimitriou2004}
{Hatzidimitriou}, D., {Antoniou}, V., {Papadakis}, I., {Kaltsa}, M.,
  {Papadaki}, C., {Papamastorakis}, I., \& {Croke}, B.~F.~W. 2004, \mnras, 348,
  1157


\bibitem[{{Ivanov} {et~al.}(2000){Ivanov}, {Borissova}, {Alonso-Herrero}, \&
  {Russeva}}]{Ivanov2000}
{Ivanov}, V.~D., {Borissova}, J., {Alonso-Herrero}, A., \& {Russeva}, T. 2000,
  \aj, 119, 2274


\bibitem[{{Jurcsik} \& {Kov{\'a}cs}(1996)}]{Jurcsik1996}
{Jurcsik}, J. \& {Kov{\'a}cs}, G. 1996, \aap, 312, 111


\bibitem[{{Kov{\'a}cs} \& {Kanbur}(1998)}]{Kovacs1998}
{Kov{\'a}cs}, G. \& {Kanbur}, S.~M. 1998, \mnras, 295, 834


\bibitem[{{Kron} \& {Guetter}(1976)}]{Kron1976}
{Kron}, G.~E. \& {Guetter}, H.~H. 1976, \aj, 81, 817


\bibitem[{{Landolt}(1992)}]{Landolt1992}
{Landolt}, A.~U. 1992, \aj, 104, 340


\bibitem[{{Lee}(1990)}]{Lee1990}
{Lee}, Y.-W. 1990, \apj, 363, 159


\bibitem[{{Massari} {et~al.}(2019){Massari}, {Koppelman}, \&
  {Helmi}}]{Massari2019}
{Massari}, D., {Koppelman}, H.~H., \& {Helmi}, A. 2019, \aap, 630, L4


\bibitem[{{Morgan} {et~al.}(2007){Morgan}, {Wahl}, \&
  {Wieckhorst}}]{Morgan2007}
{Morgan}, S.~M., {Wahl}, J.~N., \& {Wieckhorst}, R.~M. 2007, \mnras, 374, 1421


\bibitem[{{Nemec} {et~al.}(2013){Nemec}, {Cohen}, {Ripepi}, {Derekas},
  {Moskalik}, {Sesar}, {Chadid}, \& {Bruntt}}]{Nemec2013}
{Nemec}, J.~M., {Cohen}, J.~G., {Ripepi}, V., {Derekas}, A., {Moskalik}, P.,
  {Sesar}, B., {Chadid}, M., \& {Bruntt}, H. 2013, \apj, 773, 181


\bibitem[{{Piatti} \& {Carballo-Bello}(2019)}]{Piatti2019}
{Piatti}, A.~E. \& {Carballo-Bello}, J.~A. 2019, \mnras, 485, 1029


\bibitem[{{Pietrukowicz} {et~al.}(2008){Pietrukowicz}, {Olech}, {Kedzierski},
  {Zloczewski}, {Wisniewski}, \& {Mularczyk}}]{Pietrukowicz2008}
{Pietrukowicz}, P., {Olech}, A., {Kedzierski}, P., {Zloczewski}, K.,
  {Wisniewski}, M., \& {Mularczyk}, K. 2008, \actaa, 58, 121


\bibitem[{{Pols} {et~al.}(1998){Pols}, {Schr{\"o}der}, {Hurley}, {Tout}, \&
  {Eggleton}}]{Pols1998}
{Pols}, O.~R., {Schr{\"o}der}, K.-P., {Hurley}, J.~R., {Tout}, C.~A., \&
  {Eggleton}, P.~P. 1998, MNRAS, 298, 525


\bibitem[{{Pols} {et~al.}(1997){Pols}, {Tout}, {Schroder}, {Eggleton}, \&
  {Manners}}]{Pols1997}
{Pols}, O.~R., {Tout}, C.~A., {Schroder}, K.-P., {Eggleton}, P.~P., \&
  {Manners}, J. 1997, MNRAS, 289, 869


\bibitem[{{Poretti} {et~al.}(2008){Poretti}, {Clementini}, {Held}, {Greco},
  {Mateo}, {Dell'Arciprete}, {Rizzi}, {Gullieuszik}, \& {Maio}}]{Poretti2008}
{Poretti}, E., {Clementini}, G., {Held}, E.~V., {Greco}, C., {Mateo}, M.,
  {Dell'Arciprete}, L., {Rizzi}, L., {Gullieuszik}, M., \& {Maio}, M. 2008,
  \apj, 685, 947


\bibitem[{{Schlafly} \& {Finkbeiner}(2011)}]{Schlafly2011}
{Schlafly}, E.~F. \& {Finkbeiner}, D.~P. 2011, \apj, 737, 103


\bibitem[{{Schlegel} {et~al.}(1998){Schlegel}, {Finkbeiner}, \&
  {Davis}}]{Schlegel1998}
{Schlegel}, D.~J., {Finkbeiner}, D.~P., \& {Davis}, M. 1998, \apj, 500, 525


\bibitem[{{Schr{\"o}der} \& {Cuntz}(2005)}]{Schroeder2005}
{Schr{\"o}der}, K.~P. \& {Cuntz}, M. 2005, \apjl, 630, L73


\bibitem[{{Schr{\"o}der} {et~al.}(1997){Schr{\"o}der}, {Pols}, \&
  {Eggleton}}]{KPS1997}
{Schr{\"o}der}, K.-P., {Pols}, O.~R., \& {Eggleton}, P.~P. 1997, MNRAS, 285,
  696


\bibitem[{{Searle} \& {Zinn}(1978)}]{Searle1978}
{Searle}, L. \& {Zinn}, R. 1978, \apj, 225, 357


\bibitem[{{Stetson}(2000)}]{Stetson2000}
{Stetson}, P.~B. 2000, \pasp, 112, 925


\bibitem[{{Sturch}(1966)}]{Sturch1966}
{Sturch}, C. 1966, \apj, 143, 774


\bibitem[{{Taylor}(2005)}]{Taylor2005}
{Taylor}, M.~B. Astronomical Society of the Pacific Conference Series, Vol.
  347, , Astronomical Data Analysis Software and Systems XIV, ed.
  P.~{Shopbell}M.~{Britton} \& R.~{Ebert}, 29


\bibitem[{{VandenBerg} {et~al.}(2014){VandenBerg}, {Bergbusch}, {Ferguson}, \&
  {Edvardsson}}]{Vandenberg2014}
{VandenBerg}, D.~A., {Bergbusch}, P.~A., {Ferguson}, J.~W., \& {Edvardsson}, B.
  2014, \apj, 794, 72


\bibitem[{{VandenBerg} \& {Clem}(2003)}]{Vandenberg2003}
{VandenBerg}, D.~A. \& {Clem}, J.~L. 2003, \aj, 126, 778


\bibitem[{{Yepez} {et~al.}(2020){Yepez}, {Arellano Ferro}, \&
  {Deras}}]{Yepez2020}
{Yepez}, M.~A., {Arellano Ferro}, A., \& {Deras}, D. 2020, \mnras, 494, 3212


\bibitem[{Zhang {et~al.}(1996)Zhang, Ramakrishnan, \& Livny}]{Zhang1996}
Zhang, T., Ramakrishnan, R., \& Livny, M. 1996, SIGMOD Rec., 25, 103
 \href{http://doi.acm.org/10.1145/235968.233324}{\urllinklabel}

\bibitem[{{Zinn} \& {West}(1984)}]{Zinn1984}
{Zinn}, R. \& {West}, M.~J. 1984, \apjs, 55, 45


\end{thebibliography}


\end{document}